\documentstyle{article}
\normalbaselines
\newcounter{eqnletter}[equation]
\catcode`\@=11
\@addtoreset{equation}{section}   
\catcode`\@=12

\begin{document}

\begin{center}

{\LARGE\bf Infinite Quasi-Exactly Solvable Models}

\vskip 1cm

{\Large {\bf H.D. Doebner}\footnote{E-mail address: asi@pt.tu-clausthal.de}}\\

Arnold Sommerfeld Institute for Mathematical Physics,
Thechnical University of Clausthal,\\
Leibnizstrasse 10, 38-678 Clausthal, Germany\\

\vskip 0.5cm

{\Large {\bf K. {\L}azarow}}

Department of Theoretical Physics, University of {\L}\'od\'z,\\
Pomorska 149/153, 90-236 {\L}\'od\'z, Poland \\

\vskip 0.5cm

and\\

\vskip 0.5cm
 
{\Large {\bf A.G. Ushveridze}\footnote{E-mail
address: alexush@mvii.uni.lodz.pl and alexush@krysia.uni.lodz.pl} }\\

Department of Theoretical Physics, University of {\L}\'od\'z,\\
Pomorska 149/153, 90-236 {\L}\'od\'z, Poland \\

and\\

Institute of Theoretical Physics, Technical University of Clausthal,\\
Arnold Sommerfeld Str. 6, 38-678 Clausthal, Germany\\
\end{center}

\vspace{1.5cm}
\begin{abstract}

We introduce a new concept of infinite quasi-exactly
solvable models which are constructable through
multi-parameter deformations of known exactly solvable ones.
The spectral problem for these models admits exact
solutions for {\it infinitely many} eigenstates but not for the
whole spectrum. The hermiticity of their hamiltonians 
is guaranteed by construction. The proposed models have quasi-exactly
solvable classical conterparts. 

\end{abstract}

\newpage

\section{Introduction}

In this paper we present a simple method for constructing
multi-dimensional Schr\"odinger equations 
admitting exact solutions for {\it infinitely many} eigenvalues and
corresponding eigenfunctions. However, these equations
cannot be solved exactly for the whole spectrum. This fact
suggest to consider them as generalizations of known quasi-exactly
solvable problems discovered several years ago \cite{zas,turush,ush,tur}.

Remember that the distinguishing property of usual quasi-exactly solvable
problems is that the number of their explicit solutions is 
finite [5-12]. For this reason the phenomenon of quasi-exact 
solvability has a purely
quantum nature and does not admit any reasonable extension
to the classical case.  
 
A remarkable feature of models which we intend to discuss
in this paper is that they have direct {\it classical
counterparts} and the method for their construction we propose 
works equally well in both the classical and quantum cases.

In order to formulate the idea of this method it is
reasonable to start with the simpler classical case.

Let $H_0$ be a Hamilton function of an integrable classical mechanical system
whose solutions are explicitly known. Consider a
deformation of $H_0$ given by the formula
\begin{eqnarray}
H = H_0+
\sum_{i,k=1}^d(L_i-\lambda_i)U_{ik}(L_k-\lambda_k).
\label{i.1}
\end{eqnarray}
Here $L_i$ are some independent functions on the phase space being in
involution with each other and with the undeformed Hamilton
function $H_0$,
\begin{eqnarray}
\{H_0,L_i\}=0, \qquad \{L_i,L_k\}=0,
\label{i.2}
\end{eqnarray}
$\lambda_i$ are arbitrary real parameters
and  $U_{ik}$ is a matrix of arbitrarily chosen functions
which are not assumed to be in involution with $H_0$.

The classical equations of motion for the deformed Hamilton
function $H$ can be explicitly solved at least for a part of trajectories.
The proof is simple.
Consider a level surface $S$ determined by the system of
equations $L_i=\lambda_i$, $i=1,\ldots,d$. 
Because of the relations (\ref{i.2}),
$S$ is an invariant surface for the undeformed system with
Hamilton function $H_0$. Now consider the equation of motion
for the deformed system with Hamilton function $H$. It has the form
$\dot f=\{H,f\}$ or, in more detail, 
\begin{eqnarray}
\dot f = \{H_0,f\} + 
\sum_{i,k=1}^d(L_i-\lambda_i)(L_k-\lambda_k)\{U_{ik},f\}
+ 2\sum_{i,k=1}^d(L_i-\lambda_i)U_{ik}\{L_k,f\}.
\label{i.3}
\end{eqnarray}
The second and third terms in the right
hand side of (\ref{i.3}) vanish on the surfase $S$. 
This means that equations of
motion and hence their {\it solutions} for deformed and
undeformed systems coincide in the case of equal initial conditions
on $S$. Because the undeformed system has been
assumed to be integrable, the integrability of the deformed system
on the level surface $S$ is proved. But the trajectories lying on $S$
form only a part of all possible trajectories of the
deformed system. This means that the system with
Hamilton function $H$ is {\it partially integrable}. 

The extensibility of these reasonings to the quantum mechanical
case is obvious. Indeed, let $\hat H_0$ be now the hamiltonian of an 
algebraically solvable quantum mechanical system. Consider a
deformation of the hamiltonian $\hat H_0$ given by the formula
\begin{eqnarray}
\hat H = \hat H_0+
\sum_{i,k=1}^d(\hat L_i-\lambda_i)\hat U_{ik}(\hat L_k-\lambda_k).
\label{k.1}
\end{eqnarray}
Now $\hat L_i$ denote some hermitian
operators\footnote{Hereafter the hermiticity will mean the
essential self-adjoitness.} in Hilbert space commuting
with each other and with the undeformed hamiltonian $\hat H_0$,
\begin{eqnarray}
[\hat H_0,\hat L_i]=0, \qquad [\hat L_i,\hat L_k]=0,
\label{k.2}
\end{eqnarray}
$\lambda_i$ are arbitrary real parameters
and  $\hat U_{ik}$ is a symmetric matrix of arbitrarily
chosen hermitean operators
which do not generally commute with $\hat H_0$. 
The operator $\hat H$ constructed in such a
way is again a hermitean operator and thus can be viewed as a
hamiltonian for a cerain quantum mechanical model.

Let us now prove that the quantum spectral equations for
this model can be explicitly solved at least for a part of the spectrum.
The proof is essentially the same as in the classical case.
Because of the relations (\ref{k.2}), the operators $\hat
H_0$ and $\hat L_i$ admit common invariant eigensubspaces.
Denote by $\hat{\cal S}$ an invariant subspace 
corresponding to fixed eigenvalues $m_i$ of operators
$\hat L_i$. It is absolutely obvious that if 
the parameters $\lambda_i$ coincide with
these eigenvalues $m_i$, then the extra (deformation) term in
the operator $\hat H$
vanishes on all functions belonging to $\hat{\cal S}$. This means that
the spectral
problem for $\hat H$ in $\hat{\cal S}$ reduces to the 
spectral problem for $\hat H_0$ which, by assumption, is exactly
solvable. This completes the proof. 

The method described above gives an infinite series of
quasi-exactly solvable problems. 
Indeed, the phenomenon of partial solvability
appears always when the parameters $\lambda_i$
coincide with the eigenvalues $m_i$ of operators $\hat L_i$.
Another feature of these models is that for any fixed $m_i$ they form
functionally large classes which follows from 
the arbitrariness of the entries of matrix $\hat U_{ik}$ 

If we want to realize this scheme
restricting ourselves to second-order 
differential operators $\hat H_0$ and $\hat H$, then
$\hat L_i$ should be first order operators and $\hat U_{ik}$
--- ordinary functions.

In next sections of the paper we demonstrate that such models
do really exist and lead to a large variety of
multi-dimensional partially solvable quantum models defined
on compact curved manifolds. The remarkable feature of
these models is that they can be considered as models of charged
particles situated in an external electrostatic field.

\section{Some useful formulas}

We start our discussion with the investigation of transformation
properties of hermitean multi-dimensional second-order
differential operators on compact manifolds.

Let ${\cal M}$ be some smooth
compact manifold without boundaries described by 
coordinates $\xi=(\xi_1,\ldots,\xi_n)$ and characterized by
the covariant metric tensor ${\cal G}^{\mu\nu}(\xi)$. Denote by
${\cal W}$ the Hilbert space of functions on ${\cal M}$ whose
scalar product is given by the formula
\begin{eqnarray}
\langle \Psi_1,\Psi_2\rangle=\int_{\cal M} \Psi_1^*(\xi)\Psi_2(\xi)
\frac{d\xi}{\sqrt{{\cal G}(\xi)}}
\label{a.1}
\end{eqnarray}
with ${\cal G}(\xi)\equiv \det ||{\cal G}^{\mu\nu}(\xi)||$. Consider a
second order differential operator
\begin{eqnarray}
{\cal H}=-G^{\mu\nu}(\xi)\frac{\partial^2}{\partial \xi^\mu
\partial \xi^\nu}-F^\mu(\xi)\frac{\partial}{\partial \xi^\mu}-E(\xi)
\label{a.2}
\end{eqnarray}
and assume that it is hermitian in ${\cal W}$. 
Then necessarily:\\[0.2cm]

1. Functions $G^{\mu\nu}(\xi)$ are real while functions 
$F^\mu(\xi)$ and $E(\xi)$ may be complex.\\[0.2cm]

2. The following relations hold:
\begin{eqnarray}
\frac{\partial}{\partial \xi^\mu} G^{\mu\nu}(\xi)
-G^{\mu\nu}(\xi)\frac{\partial}{\partial \xi^\mu}\ln
\sqrt{{\cal G}(\xi)}=
\mbox{Re}\ F^\nu(\xi), 
\label{a.3}
\end{eqnarray}
\begin{eqnarray}
\frac{\partial}{\partial \xi^\mu} \mbox{Im}\ F^{\mu}(\xi)
-\mbox{Im}\ F^{\mu}(\xi)\frac{\partial}{\partial \xi^\mu}\ln 
\sqrt{{\cal G}(\xi)}=
2\mbox{Im}\ E(\xi). 
\label{a.4}
\end{eqnarray}
If functions $F^\mu(\xi)$ and $E(\xi)$ are real then the
second relation disappears.  \\[0.2cm]

3. If the matrix $G^{\mu\nu}(\xi)$ is positive definite so
that $G(\xi)=\det ||G^{\mu\nu}(\xi)||$ is a positive function
on ${\cal M}$ then there exists a homogeneous transformation
\begin{eqnarray}
H=\Phi^{-1}{\cal H}\Phi
\label{a.5}
\end{eqnarray}
realized by real function 
\begin{eqnarray}
\Phi=\left(\frac{{\cal G}(\xi)}{G(\xi)}\right)^{1/4} 
\label{a.6}
\end{eqnarray}
which transforms the operator ${\cal H}$ into the form
\begin{eqnarray}
H= -\sqrt{G(\xi)}\left(
\frac{\partial}{\partial \xi^\mu} -\imath A_\mu(\xi)\right)
\left\{\frac{G^{\mu\nu}(\xi)}{\sqrt{G(\xi)}}\left(
\frac{\partial}{\partial \xi^\nu} -\imath A_\nu(\xi)\right)\right\}
+V(\xi).
\label{a.7}
\end{eqnarray}
Here 
\begin{eqnarray}
A_{\mu}(\xi)= \frac{1}{2}G_{\mu\nu}(\xi) \mbox{Im}\ F^\nu(\xi) 
\label{a.8}
\end{eqnarray}
and 
\begin{eqnarray}
V(\xi)&=&-G^{\mu\nu}(\xi)\left[
A_\mu(\xi)A_\nu(\xi)+\Phi^{-1}(\xi)\frac{\partial^2}{\partial
\xi^\mu \partial \xi^\nu}\Phi(\xi) \right]\nonumber\\
&&-\mbox{Re}\ F^\mu(\xi)\Phi^{-1}(\xi)\frac{\partial}{\partial
\xi^\mu}\Phi(\xi)- \mbox{Re}\ E(\xi)  
\label{a.9}
\end{eqnarray}
The operator $H$ can be interpreted as hamiltonian of a charged
quantum particle moving on a certain curved manifold $M$
(which is described by the covariant metric tensor
$G^{\mu\nu}(\xi)$), interacting with electrostatic field
$A_{\mu}(\xi)$ and with external potential $V(\xi)$.

\section{The undeformed algebraically solvable models}

Let $\Gamma$ be a compact finite-dimensional Lie group and
$\Lambda$ be the corresponding Lie algebra. The Weyl generators of
algebra $\Lambda$ associated with its simple roots $\pi_i$ and
positive roots $\alpha$ we denote by $L_i^0$ and
$L_\alpha^\pm$, respectively. We have:
\begin{eqnarray}
[L_i^0,L_k^0]=0,  \qquad [L_i^0, L_\alpha^\pm]=\pm (\pi_i,\alpha)L_\alpha^\pm
\label{b.1}
\end{eqnarray}
It is known that these generators can be realized as vector
fields on the homogeneous space
\begin{eqnarray}
{\cal M}=\Gamma/\Gamma_0,
\label{b.2}
\end{eqnarray}
where $\Gamma_0$ is a certain stationary subgroup of $\Gamma$.
Introducing the coordinates
$\xi^\mu$ on ${\cal M}$, we can write 
\begin{eqnarray}
L_i^0=T_a^{0\mu}(\xi)\frac{\partial}{\partial \xi^\mu},
\qquad 
L_\alpha^\pm=T_a^{\pm\mu}(\xi)\frac{\partial}{\partial \xi^\mu}
\label{b.3}
\end{eqnarray}

Now note that the compactness of $\Gamma$ implies the
compactness of ${\cal M}$.
This enables one to introduce on ${\cal M}$ a $\Gamma$-invariant metric
${\cal G}^{\mu\nu}(\xi)$ with elements given by 
the following simple formula: 
\begin{eqnarray}
{\cal G}^{\mu\nu}(\xi)= K^{ik} T_i^{0\mu}(\xi)T_k^{0\nu}(\xi)+
\sum_\alpha\left[T_\alpha^{+\mu}(\xi)T_\alpha^{-\nu}(\xi)+
T_\alpha^{-\mu}(\xi)T_\alpha^{+\nu}(\xi)\right].
\label{b.4}
\end{eqnarray}
in which $K^{ik}$ is the Cartan part of the Killing tensor. 
It is known that the neutral generators $L_i^0$ are anti-hermitean
with respect to the scalar product (\ref{a.1}) with metric (\ref{b.4}),
\begin{eqnarray}
(L_i^0)^+ = -L_i^0,  
\label{b.5}
\end{eqnarray}
while the raising and lowering generators $L_\alpha^\pm$
are anti-hermitian conjugated
\begin{eqnarray}
(L_\alpha^\pm)^+ = -L_\alpha^\mp.  
\label{b.6}
\end{eqnarray}

Consider the operator
\begin{eqnarray}
{\cal H}_0=\sum h^{ik}L_i^0 L_k^0 +\sum f_\alpha^+ L_\alpha^+L_\alpha^-
+\sum f_\alpha^- L_\alpha^-L_\alpha^+
\label{b.7}
\end{eqnarray}
in which $h^{ik}$ and $f_\alpha^\pm$ are arbitrary real constants.
Obviously, this operator is hermitean (by construction)
with respect to the metric (\ref{a.1}) and (\ref{b.4}),
\begin{eqnarray}
{\cal H}_0^+={\cal H}_0,
\label{b.8}
\end{eqnarray}
moreover, it is a second order differential operator of the form (\ref{a.2}),
and therefore, according to the theorem of the previous
section, it always can be reduced to the form (\ref{a.7}).

Let us now show that the obtained quantum model is
exactly solvable. This means that the whole spectrum of its hamiltonian
can be obtained algebraically. In order to demonstrate this fact, 
let us repeat the reasonings of ref. [7] and 
consider the quadratic Casimir operator for the group $\Gamma$: 
\begin{eqnarray}
\Delta = K^{ik} L_i^{0}L_k^{0}+ \sum_\alpha\left[L_\alpha^{+}L_\alpha^{-}+
L_\alpha^{-}L_\alpha^{+}\right].
\label{b.9}
\end{eqnarray}
This operator, which can be interpreted as the
Laplace operator on ${\cal M}$, is hermitian, non-degenerate and commutes
with the hamiltonian ${\cal H}$. From this it follows that the
Hilbert space in which the hamiltonian $H$ acts splits into a
sum of eigensubspaces of the operator $\Delta$. Since $\Delta$ 
commutes with generators $L_i^0$ and $L_\alpha^\pm$, 
every such subspace forms a finite-dimensional
representation of the group $\Gamma$. The basis functions in
these representation spaces are the so-called ``generalized
spherical harmonics'' whose concrete form can be
constructed in local coordinates $\xi$ are fixed. 
In the basis of the generalized
spherical harmonics the hamiltonian ${\cal H}$ takes the block
diagonal form. Each block has a finite dimension and is
completely disconnected from all others. 
Therefore, the spectral problem for this hamiltonian breaks
up into an infinite number of finite-dimensional spectral problems,
each of which can be solved algebraically. 

\section{The deformed partially solvable model}

Let us now construct a new (deformed) operator ${\cal H}$
given by the formula
\begin{eqnarray}
{\cal H} = {\cal H}_0 + 
\sum_{i,k}(L_i^0-\imath\lambda_i)U_{ik}(L_k^0-\imath\lambda_k)
\label{e.1}
\end{eqnarray}
in which $U_{ik}$ is an arbitrary real symmetric matrix whose
entries are arbitrarily chosen 
smooth functions on the manifold ${\cal M}$ and
$\lambda_i$ are some real parameters. Note that the
deformed operator $\bar H$ is again hermitean
\begin{eqnarray}
{\cal H}^+={\cal H},
\label{e.2}
\end{eqnarray}
it is again a second order operator and thus can be reduced
to the form (\ref{a.7}).

Now, however, we cannot claim any longer that this
operator admits a complete algebraization of the
spectral problem as it was in the case of ${\cal H}_0$. Indeed, the
reasonings given in the previous section for ${\cal H}_0$ do not
work because the operator ${\cal H}$ does not commute with
the Laplace operator $\Delta$ (i.e. with the second-order
Casimir invariant of the group $\Gamma$).
The reason for this is the presence of the extra
deformation term containing the arbitrarily chosen
functions $U_{ik}$.

Nevertheless, from the fact of commutativity of
${\cal H}_0$ with generators $L_i^0$ and 
the reasonings given in section 1 it follows
that the operator ${\cal H}$ for some
quantized values of parameters $\lambda_i=m_i$ (i.e coinciding with
eigenvalues of operators $L_i^0$) admits a
partial algebraization of the spectral problem in the sense
that one can construct purely algebraically a certain
set of its eigenvalues and eigenfunctions.
The number of these solutions is obviously equal to the
dimension of the corresponding common eigensubspace of
operators $L_i^0$. But this dimension is equal to infinity
because the number of eigenstates of operators $L_i^0$
having one and the same projections $m_i$ of the 
"generalized angular momenta" is infinite.

This reasoning completes the exposition of our method for
building partially solvable models in an external
electrostatic field.

\section{An example}

Following general prescriptions of section 3, let us
consider the case of algebra $so(3)$ with three  generators
$L^0$, $L^+$, $L^-$ obeying the commutation relations
\begin{eqnarray}
[L^+,L^-]=2 L^0,  \qquad [L^0, L^\pm]=\pm  L^\pm
\label{f.2}
\end{eqnarray}
Choosing the operator (\ref{b.7}) in the form
\begin{eqnarray}
{\cal H}_0=a(L^0)^2 +b( L^+L^-
+L^-L^+)+cL^0
\label{f.1}
\end{eqnarray}
and rewriting the generators in the spherical coordinates as
\begin{eqnarray}
L^{+}=(\cos\phi+i\sin\phi)\frac{\partial}{\partial\theta}+
\mbox{ctg}\theta(i\cos\phi-\sin\phi)\frac{\partial}{\partial\phi}
\label{f.3}
\end{eqnarray}
\begin{eqnarray}
L^{-}=(-\cos\phi+i\sin\phi)\frac{\partial}{\partial\theta}+
\mbox{ctg}\theta(i\cos\phi+\sin\phi)\frac{\partial}{\partial\phi}
\label{f.4}
\end{eqnarray}
\begin{eqnarray}
L^0=-i\frac{\partial}{\partial\phi}
\label{f.5}
\end{eqnarray}
we obtain for ${\cal H}_0$:
\begin{eqnarray}
{\cal H}_0=-2b\frac{\partial^2}{\partial\theta^2}-
(a+2\mbox{ctg}^2\theta)\frac{\partial^2}{\partial\phi^2}-
2b\mbox{ctg}\theta\frac{\partial}{\partial\theta}-
ic\frac{\partial}{\partial\phi}
\label{f.6}
\end{eqnarray}

This model is algebraically solvable because it commutes
with Laplasian (see general discussion in section 3).
Let us now construct the deformed operator ${\cal H}$ 
given by formula (\ref{e.1}).
We know that, in this case, we can not expect any longer a 
complete algebraization 
of the spectral problem, but if $\lambda$ coincides 
with eigenvalues $m$ of the operator 
$L^0$, a partial algebraization will be realized.
Let $\lambda=m$ where $m$ satisfies the equation $L^0\Psi=m\Psi$.
In this case ${\cal H}$ takes the form:
\begin{eqnarray}
{\cal H}=(-a-2\mbox{ctg}^2\theta+U(\theta,\phi) )
\frac{\partial ^2}{\partial\phi ^2}
-2b\frac{\partial ^2}{\partial \theta ^2}-2b\mbox{ctg}\theta 
\frac {\partial}{\partial \theta }\nonumber\\
 -(- \imath c+\frac{\partial U(\theta ,\phi )}{\partial \phi }-2 
\imath mU(\theta ,\phi))\frac \partial {\partial \phi }-
(-\imath m \frac{\partial U(\theta ,\phi )}{
\partial \phi }-m^2 U(\theta ,\phi ))
\label{f.7}
\end{eqnarray}
where $U(\theta,\phi)$ is an arbitrarily fixed real function.

Operator ${\cal H}$ is hermitian in a space of suitable 
functions on the
sphere and is endowed with a scalar product given by
\begin{eqnarray}
\langle \Psi_1,\Psi_2\rangle=
\int_{S^2} \Psi_1^*(\theta,\phi)\Psi_2(\theta,\phi)
\sin \theta d\theta d\phi.
\label{f.8}
\end{eqnarray}
According to the results of section 2, (\ref{f.7}) 
can be represented in the form (\ref{a.7}) 
with $\xi_1 \rightarrow \theta$,
$\xi_2 \rightarrow \phi$ and with
\begin{eqnarray}
G^{11}=2b,\quad
G^{22}=a+2\mbox{ctg}^2\theta+U(\theta,\phi),\quad
G^{12}=G^{21}=0,
\end{eqnarray}
\begin{eqnarray}
G=\det G^{\mu \nu}=2b\left( a+2\mbox{ctg}^2\theta+U(\theta,\phi) \right). 
\label{f.10}
\end{eqnarray}
Expanding (\ref{a.7}) and comparing the obtained formula with (\ref{f.7}), 
we find the components of the external field $A_\mu$,
\begin{eqnarray}
A_{1}(\theta,\phi)=0
&&A_{2}(\theta,\phi)=\frac{c-2mU(\theta,\phi)}{2\left( a+
2\mbox{ctg}^2\theta+U(\theta,\phi) \right) }, 
\label{f.11}
\end{eqnarray}
and the potential of the model
\begin{eqnarray}
V\left( \theta ,\phi \right)=-\frac{\left( c-2mU\left( \theta ,\phi
\right) \right) ^2}{4\left( a+2\mbox{ctg}^2\theta +
U\left( \theta ,\phi \right)
\right) }+m^2U\left( \theta ,\phi \right)\nonumber\\
+\frac{2b\mbox{ctg}\theta \left( \left( a-2+U\left( \theta ,\phi \right) 
\right)
\sin 2\theta +\sin ^2\theta \frac{\partial U\left( \theta ,\phi \right) }{%
\partial \theta }\right) +\sin ^2\theta \left( \frac{\partial U\left( \theta
,\phi \right) }{\partial \phi }\right) ^2}{4\sqrt{a\sin ^2\theta +2\cos
^2\theta +\sin ^2\theta \cdot U\left( \theta ,\phi \right) }}\nonumber\\
-\frac{3\left( \frac{\partial U\left( \theta ,\phi \right) }{\partial \phi }%
\right) ^2}{16\sqrt{a\sin ^2\theta +2\cos ^2\theta +\sin ^2\theta  
\cdot U\left(
\theta ,\phi \right) }}\nonumber\\
+\frac 1{4\sqrt{a\sin ^2\theta +2\cos ^2\theta +\sin ^2\theta \cdot U\left(
\theta ,\phi \right) }}\frac{\partial ^2U\left( \theta ,\phi \right) }{%
\partial \phi ^2}\nonumber\\
-\frac{6b\left( \left( a-2+U\left( \theta ,\phi \right) \right) \sin
2\theta +\sin ^2\theta \frac{\partial U\left( \theta ,\phi \right) }{%
\partial \theta }\right) ^2}{16\left( a\sin ^2\theta +2\cos ^2\theta +\sin
^2\theta \cdot U\left( \theta ,\phi \right) \right) ^{\frac 32}}\nonumber\\
+\frac{2b\left( 2\frac{\partial U\left( \theta ,\phi \right) }{\partial
\theta }\sin 2\theta +2\left( a-2+U\left( \theta ,\phi \right) \right) 
\cos 2\theta
+\sin ^2\theta \frac{\partial ^2U\left( \theta ,\phi \right) }{%
\partial \theta ^2}\right) }{\sqrt{a\sin ^2\theta +2\cos ^2\theta +\sin
^2\theta \cdot U\left( \theta ,\phi \right) }}
\label{f.12}
\end{eqnarray}
Formulas (\ref{f.11}) and (\ref{f.12}) complete the construction 
of quasi-exactly solvable models with hermitian
hamiltonians having an infinite number of exact solutions.

\end{document}